\documentclass[twocolumn,prl,showpacs,preprintnumbers,amsmath,amssymb]{revtex4}
\usepackage[T1]{fontenc}
\usepackage[latin1]{inputenc}
\usepackage{amsmath}
\usepackage{graphicx}
\usepackage{amssymb}

\makeatletter

\usepackage{epsfig}

\usepackage{dcolumn}
\usepackage{bm}
\usepackage{wasysym}

\usepackage{marvosym}

\usepackage{times}

\usepackage{color}

\makeatletter

\usepackage{ulem}

\makeatother

\makeatother
\begin{document}

\title{Localized states at zigzag edges of bilayer graphene}

\author{Eduardo V. Castro$^{1}$, N. M. R. Peres$^{2}$, J. M. B. Lopes dos
Santos$^{1}$, A. H. Castro Neto$^{3}$, and F. Guinea$^{4}$}

\affiliation{$^{1}$ CFP and Departamento de F\'{\i}sica, Faculdade de Ci\^{e}ncias
Universidade do Porto, P-4169-007 Porto, Portugal}

\affiliation{$^{2}$Center of Physics and Departamento de F\'{\i}sica, Universidade
do Minho, P-4710-057 Braga, Portugal}

\affiliation{$^{3}$Department of Physics, Boston University, 590 Commonwealth
Avenue, Boston, MA 02215, USA}

\affiliation{$^{4}$Instituto de Ciencia de Materiales de Madrid. CSIC. Cantoblanco.
E-28049 Madrid, Spain}

\begin{abstract}
We report the existence of zero energy surface states localized at
zigzag edges of bilayer graphene. Working within the tight-binding
approximation we derive the analytic solution for the wavefunctions
of these peculiar surface states. It is shown that zero energy edge
states in bilayer graphene can be divided into two families: (\emph{i}) states
living only on a single plane, equivalent to surface states in monolayer
graphene; (\emph{ii}) states with finite amplitude over the two layers,
with an enhanced penetration into the bulk. The bulk and surface (edge)
electronic structure of bilayer graphene nanoribbons is also studied,
both in the absence and in the presence of a bias voltage between
planes. 
\end{abstract}


\maketitle

%

\emph{Introduction:} The quest for new materials and material properties
has recently led to graphene, the missing two-dimensional (2D) allotrope
of carbon \cite{NGM+04}. Stability and ballistic transport on the
submicrometre scale, even at room-temperature, make graphene based
electronics a promising possibility \cite{GN07}. Indeed, with Si-based
technology approaching its limits, a truly 2D material with unconventional
electronic properties is regarded with great expectations. 

Graphene is a zero-gap semiconductor, and this prevents standard logic
applications where the presence of a finite gap is paramount. Band
gaps can still be engineered by confining graphene electrons in narrow
ribbons \cite{CLR+07,HOZ+07}. However, the lateral confinement brings
about the presence of edges, which in graphene can have profound consequences
on electronics. This is essentially due to the rather different behavior
of the two possible (perfect) terminations in graphene: \emph{zigzag}
and \emph{armchair}. While zigzag edges support localized states,
armchair edges do not \cite{japonese,Dresselhaus,WFA+99}. These edge
states occur at zero energy, the same as the Fermi level of undoped
graphene, meaning that low energy properties may be substantially
altered by their presence. The self-doping phenomenon \cite{PGN06}
and the edge magnetization with consequent gap opening \cite{SCLprl06}
are among edge states driven effects.

Bilayer graphene, as its single single layer counterpart, is also
a zero gap semiconductor \cite{MF06}, but only in the absence of
an external electric field: the electronic gap can be tuned externally
\cite{ENM+06}. Nevertheless, the question regarding the presence
of edge states in bilayer graphene is pertinent. Firstly, zigzag edges
are among the possible terminations in bilayer graphene, and secondly,
the presence of edges is unavoidable in tiny devices.

\begin{figure}[t]
\begin{centering}
\includegraphics[width=0.95\columnwidth]{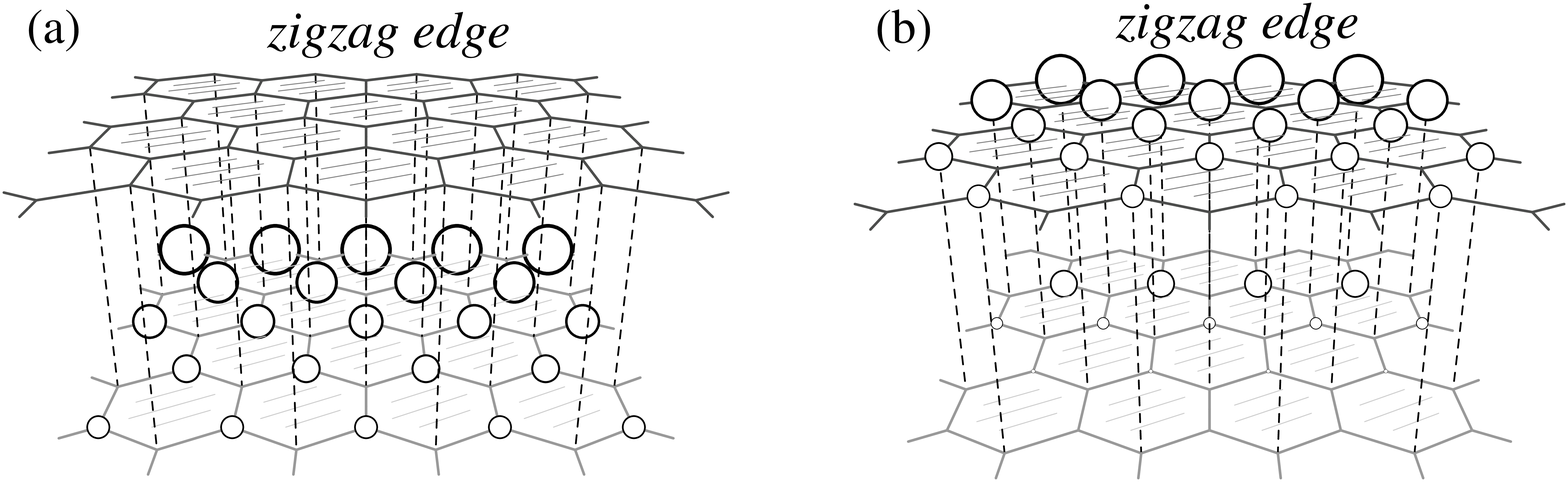}
\par\end{centering}

\caption{\label{fig:esLattice}Charge density representation for the two families
of edge states in bilayer graphene at $ka/2\pi=0.35$: \emph{monolayer}~(a),
given by Eq.~(\ref{eq:solB}); and \emph{bilayer}~(b), given by
Eq.~(\ref{eq:solAn}).}
\end{figure}

In the present paper we show that zero energy edge states do exist
at zigzag edges of bilayer graphene. An analytic solution for the
wavefunction is given assuming a semi-infinite system and a first
nearest-neighbor tight-binding model. The analytic solution we have
found defines two types of edge states: monolayer edge states, with
finite amplitude on a single plane; and bilayer edge states, with
finite amplitude on both planes, and with an enhanced penetration
into the bulk. A schematic representation of the two families of edge
states can be seen in Fig.~\ref{fig:esLattice}. We also show that
bilayer graphene nanoribbons with zigzag edges have four flat bands
occurring at zero energy, consequence of the two families of edge
states localized on each edge. In the case of a biased ribbon, where
the two planes are at different electrostatic potential and a band
gap develops for the bulk electronic structure, the spectrum still
shows two flat bands while the other two give rise to level crossing
inside the gap.

%

\begin{figure}[t]
\begin{centering}
\includegraphics[width=1\columnwidth,width=0.9\columnwidth]{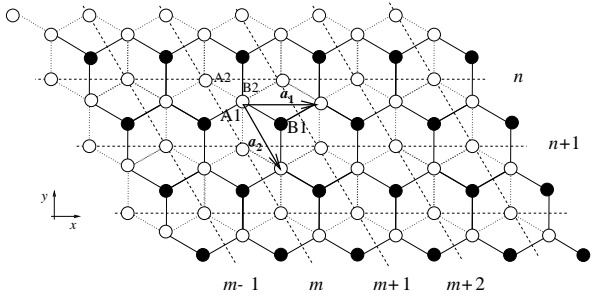}
\par\end{centering}

\caption{\label{cap:ribbon}Ribbon geometry with zigzag edges for bilayer
graphene.}
\end{figure}

\emph{Surface states in semi-infinite bilayer graphene:} The study
of edge states in $AB-$stacked bilayer graphene given here is based
on the ribbon geometry with zigzag edges shown in Fig.~\ref{cap:ribbon}.
We use labels~1 and~2 for the top and the bottom layers, respectively,
and labels $Ai$ and $Bi$ for each of the two sublattices in layer~$i$.
Each four-atom unit cell (parallelograms in Fig.~\ref{cap:ribbon})
has integer indices $m$~(longitudinal) and $n$~(transverse) such
that $m\mathbf{a}_{1}+n\mathbf{a}_{2}$ is its position vector, where
$\mathbf{a}_{1}=a(1,0)$ and $\mathbf{a}_{2}=a(1,-\sqrt{3})/2$ are
the basis vectors and $a\approx2.46\,\textrm{\AA}$ is the lattice
constant. The simplest model one can write to describe non-interacting
electrons in $AB$-stacked bilayer is the first nearest-neighbor tight-binding
model given by,%
%
\begin{multline}
H=-t\sum_{i=1}^{2}\sum_{m,n}a_{i;m,n}^{\dagger}(b_{i;m,n}+b_{i;m-1,n}+b_{i;m,n-1})\\
-t_{\perp}\sum_{m,n}a_{1;m,n}^{\dagger}b_{2;m,n}+\textrm{h.c.},\label{eq:H}\end{multline}
where $a_{i;m,n}$ ($b_{i;m,n}$) is the annihilation operator for
the state in sublattice $Ai$ ($Bi$), $i=1,2$, at position ($m,n$).
The first term in Eq.~(\ref{eq:H}) describes in-plane hopping while
the second term parametrizes the inter-layer coupling ($t_{\perp}/t\ll1$).
Without loss of generality we assume that the ribbon has $N$ unit
cells in the transverse cross section ($y$ direction) with $n\in\{0,\dots,N-1\}$,
and we use periodic boundary conditions along the longitudinal direction
($x$ direction). This last simplification enables the diagonalization
of Hamiltonian~(\ref{eq:H}) with respect to the $m$~index just
by Fourier transform along the longitudinal direction, $H=\sum_{k}\, H_{k}$,
with $H_{k}$ given by,%
%
\begin{multline}
H_{k}=-t\sum_{i=1}^{2}\sum_{n}a_{i;k,n}^{\dagger}[(1+e^{ika})b_{i;k,n}+b_{i;k,n-1}]\\
-t_{\perp}\sum_{n}a_{1;k,n}^{\dagger}b_{2;k,n}+\textrm{h.c.}\,.\label{eq:Hk}\end{multline}

\begin{figure}[t]
\begin{centering}
\includegraphics[scale=0.37,width=0.9\columnwidth]{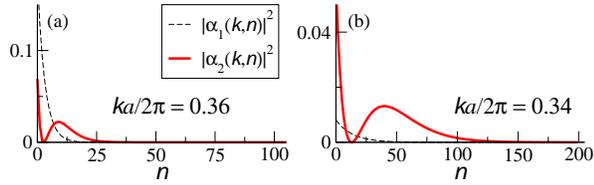}
\par\end{centering}

\caption{\label{cap:edgestates}(Color online) (a)~-~Charge density for
bilayer edge states at $ka/2\pi=0.36$. (b)~-~The same as in (a)
at $ka/2\pi=0.34$.}
\end{figure}

In order to search for zero-energy edge states we solve the Schrödinger
equation, $H_{k}\left|\mu,k\right\rangle =E_{\mu,k}\left|\mu,k\right\rangle $,
for $E_{\mu,k}=0$. First we note that Hamiltonian $H_{k}$ in Eq.~(\ref{eq:Hk})
effectively defines a 1D problem in the transverse direction of the
ribbon. It is then possible to write any eigenstate $\left|\mu,k\right\rangle $
as a linear combination of the site amplitudes along the cross section,%
%
 \begin{equation}
\left|\mu,k\right\rangle =\sum_{n}\sum_{i=1}^{2}\big[\alpha_{i}(k,n)\left|a_{i},k,n\right\rangle +\beta_{i}(k,n)\left|b_{i},k,n\right\rangle \big],\label{eq:eigstate}\end{equation}
where the four terms per~$n$ refer to the four atoms per unit cell,
to which we associate the one-particle states $\left|c_{i},k,n\right\rangle =c_{i;k,n}^{\dagger}\left|0\right\rangle $,
with $c_{i}=a_{i},b_{i}$, and $i=1,2$. In addition we require the
following boundary conditions,%
%
\begin{equation}
\alpha_{1}(k,N)=\alpha_{2}(k,N)=\beta_{1}(k,-1)=\beta_{2}(k,-1)=0,\label{eq:BC}\end{equation}
accounting for the finite width of the ribbon. Then, analogously to
the single layer case, it can be straightforwardly shown that if Eq.~(\ref{eq:eigstate})
is a zero energy solution of the Schrödinger equation, the coefficients
satisfy the following matrix equations:%
%
 \begin{align}
\begin{bmatrix}\alpha_{1}(k,n+1)\\
\alpha_{2}(k,n+1)\end{bmatrix} & =e^{-i\frac{ka}{2}}\begin{bmatrix}D_{k} & 0\\
-\frac{t_{\perp}}{t}e^{i\frac{ka}{2}} & D_{k}\end{bmatrix}\begin{bmatrix}\alpha_{1}(k,n)\\
\alpha_{2}(k,n)\end{bmatrix},\label{eq:matrixA}\\
\begin{bmatrix}\beta_{2}(k,n-1)\\
\beta_{1}(k,n-1)\end{bmatrix} & =e^{i\frac{ka}{2}}\begin{bmatrix}D_{k} & 0\\
-\frac{t_{\perp}}{t}e^{-i\frac{ka}{2}} & D_{k}\end{bmatrix}\begin{bmatrix}\beta_{2}(k,n)\\
\beta_{1}(k,n)\end{bmatrix},\label{eq:matrixB}\end{align}
where $D_{k}=-2\cos(ka/2)$. As the $2\times2$ matrix in Eqs.~(\ref{eq:matrixA})
and~(\ref{eq:matrixB}) has the following property for any complex
$p_{k}$,%
%
 \begin{equation}
\begin{bmatrix}D_{k} & 0\\
p_{k} & D_{k}\end{bmatrix}^{n}=\begin{bmatrix}D_{k}^{n} & 0\\
nD_{k}^{n-1}p_{k} & D_{k}^{n}\end{bmatrix},\label{eq:mp}\end{equation}
 we conclude by induction that the general solution of Eqs.~(\ref{eq:matrixA})
and~(\ref{eq:matrixB}) has the form:%
%
 \begin{align}
\begin{bmatrix}\alpha_{1}(k,n)\\
\alpha_{2}(k,n)\end{bmatrix} & =e^{-i\frac{ka}{2}n}\,\mathbf{T}_{n}\begin{bmatrix}\alpha_{1}(k,0)\\
\alpha_{2}(k,0)\end{bmatrix},\label{eq:gmatrixA}\\
\begin{bmatrix}\beta_{2}(k,N-n-1)\\
\beta_{1}(k,N-n-1)\end{bmatrix} & =e^{i\frac{ka}{2}n}\,\mathbf{T}_{n}^{*}\begin{bmatrix}\beta_{2}(k,N-1)\\
\beta_{1}(k,N-1)\end{bmatrix},\label{eq:gmatrixB}\end{align}
 for $n\geq1$, where the matrix $\mathbf{T}_{n}$ is given by,%
%
 \begin{equation}
\mathbf{T}_{n}=\begin{bmatrix}D_{k}^{n} & 0\\
-nD_{k}^{n-1}\frac{t_{\perp}}{t}e^{i\frac{ka}{2}} & D_{k}^{n}\end{bmatrix},\label{Tn}\end{equation}
 and $\mathbf{T}_{n}^{*}$ is the matrix whose elements are the complex
conjugate of $\mathbf{T}_{n}$. One also requires the convergence
condition $\left|-2\cos(ka/2)\right|<1$, which guarantees that Eq.~(\ref{eq:BC})
is satisfied for semi-infinite systems. It is easy to see that the
semi-infinite bilayer sheet has edge states for $k$ in the region
$2\pi/3<ka<4\pi/3$, which corresponds to $1/3$ of the possible $k$'s,
as in the graphene sheet. The next question concerns the number of
edge states. As any initialization vector is a linear combination
of only two linearly independent vectors there are only two states
per edge (per $k$). Moreover, Eqs~(\ref{eq:gmatrixA}) and~(\ref{eq:gmatrixB})
are edge states solutions on different sides of the ribbon. When the
semi-infinite system is considered only one of them survives. In particular,
taking the limit $N\rightarrow\infty$, and choosing the simplest
linear independent initialization vectors $[\alpha_{1}(k,0),0]$ and
$[0,\alpha_{2}(k,0)]$, the two possible edge states are,%
%
 \begin{align}
 & \begin{cases}
\alpha_{1}(k,n)=\alpha_{1}(k,0)D_{k}^{n}e^{-i\frac{ka}{2}n}\\
\alpha_{2}(k,n)=-\alpha_{1}(k,0)nD_{k}^{n-1}\frac{t_{\perp}}{t}e^{-i\frac{ka}{2}(n-1)}\end{cases},\label{eq:solA}\\
\intertext{and} & \begin{cases}
\alpha_{1}(k,n)=0\\
\alpha_{2}(k,n)=\alpha_{2}(k,0)D_{k}^{n}e^{-i\frac{ka}{2}n}\end{cases}.\label{eq:solB}\end{align}
Although linearly independent, it is clear that the edge states~(\ref{eq:solA})
and~(\ref{eq:solB}) are not orthogonal, except for $ka=\pi$. It
is convenient to orthogonalize~(\ref{eq:solA}) with respect to~(\ref{eq:solB})
so that we finally obtain,%
%
\begin{equation}
\begin{cases}
\alpha_{1}(k,n)=\alpha_{1}(k,0)D_{k}^{n}e^{-i\frac{ka}{2}n}\\
\alpha_{2}(k,n)=-\alpha_{1}(k,0)D_{k}^{n-1}\frac{t_{\perp}}{t}e^{-i\frac{ka}{2}(n-1)}\Big(n-\frac{D_{k}^{2}}{1-D_{k}^{2}}\Big)\end{cases},\label{eq:solAn}\end{equation}
which, together with Eq.~(\ref{eq:solB}), represent all possible
orthonormalized zero-energy edge states for bilayer graphene. The
normalization constants in Eqs.~(\ref{eq:solB}) and~(\ref{eq:solAn})
are given by,%
%
 \begin{align}
|\alpha_{1}(k,0)|^{2} & =\frac{(1-D_{k}^{2})^{3}}{(1-D_{k}^{2})^{2}+t_{\perp}^{2}/t^{2}},\label{eq:normconstA}\\
|\alpha_{2}(k,0)|^{2} & =1-D_{k}^{2}.\label{eq:normconstB}\end{align}
An example of the charge density associated with Eq.~(\ref{eq:solAn})
is shown in Fig.~(\ref{cap:edgestates}) for $t_{\perp}/t=0.2$,
where the $|\alpha_{1}(k,n)|^{2}$ dependence can also be seen as
the solution given by Eq.~(\ref{eq:solB}) for $|\alpha_{2}(k,n)|^{2}$,
apart from a normalization factor.

The solution given by Eq.~(\ref{eq:solB}) is exactly the same as
that found for a single graphene layer \cite{japonese,Dresselhaus},
where the only sites with non-zero amplitude belong to the~$A$ sublattice
of layer~2, the one disconnected from the other layer. Solution~(\ref{eq:solAn}),
on the other hand, is an edge state that can only be found in bilayer
graphene. The sites of non-vanishing amplitude for this edge state
occur at sublattice~$A$ of layer~2, and at sublattice~$A$ of
layer~1, which is connected to the other layer through $t_{\perp}$
(see Fig.~\ref{fig:esLattice}). Had we increased the semi-infinite
sheet from the other side of the ribbon, and two similar edge states
would have appeared in the opposite edge with non-zero amplitudes
at sites of the~$B$ sublattices. In regards to the penetration depth,
$\lambda$, it is easily seen from Eqs.~(\ref{eq:solB}) and~(\ref{eq:solAn})
that both solutions have the same value: $\lambda=-1/\ln|D_{k}|$.
Nevertheless, the solution given by Eq.~(\ref{eq:solAn}) has a linear
dependence in $n$ which enhances its penetration into the bulk. We
expect these states to contribute more to self doping then the usual
single layer edge states \cite{PGN06}, as the induced Hartree potential
which limits the charge transfer between the bulk and the edge will
be weaker. Note that the key to self doping is the presence of both
an electron-hole asymmetry and extended defects. Electron-hole asymmetry
may be due to in-plane next nearest-neighbor hopping (NNN) $t'$,
while edges play the role of extended defects. The finite $t'$ shifts
the energy of edge states, leading to charge transfer between clean
regions and defects. The energy shift for the single layer is given
by $E_{k}\approx-t'(D_{k}^{2}-1)$ to first order in $t'$, apart
from a global factor of $-3t'$ \cite{SMS06}. This is exactly the
energy shift we get (away from the Dirac points) for bilayer graphene
with in-plane NNN hopping, if we neglect terms of the order $t't_{\perp}/t$
and higher.

%

\begin{figure}[t]
\begin{centering}
\includegraphics[scale=0.37,width=0.9\columnwidth]{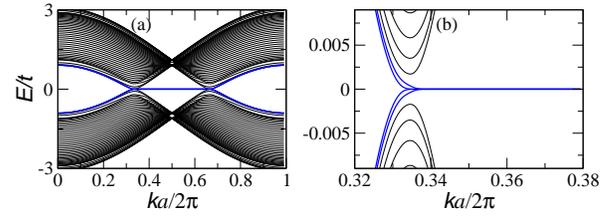}
\par\end{centering}

\caption{\label{cap:edgestateRibbon}(Color online) (a)~-~Energy spectrum
for a graphene bilayer ribbon with zigzag edges for $N=400$. (b)~-~Zoom
of panel~(a). The inter-layer coupling was set to $t_{\perp}/t=0.2$.}
\end{figure}

\emph{Nanoribbons of bilayer graphene (unbiased):} So far we studied
localized states at the semi-infinite bilayer graphene. Experimentally,
however, the relevant situation is a bilayer ribbon. The band structure
of a bilayer ribbon with zigzag edges is shown in Fig.~\ref{cap:edgestateRibbon}~(a)
for $N=400$, obtained by numerically solving Eq.~(\ref{eq:Hk}).
We can see the partly flat bands at $E=0$ for $k$ in the range $2\pi/3\leq ka\leq4\pi/3$,
corresponding to four edge states, two per edge. The zoom shown in
Fig.~\ref{cap:edgestateRibbon}~(b) for $ka\approx2\pi/3$ clearly
shows that there are four flat bands. Strictly speaking, the edge
states given by Eqs.~(\ref{eq:solB}) and~(\ref{eq:solAn}) {[}and
the other two resulting from Eq.~(\ref{eq:gmatrixB})] are eigenstates
of the semi-infinite system only. In the ribbon the overlapping of
four edge states leads to a slight dispersion and non-degeneracy.
However, as long as the ribbon width is sufficiently large, this effect
is only important at $ka\simeq2\pi/3$ and $ka\simeq4\pi/3$ where
$\lambda$ is large enough for the overlapping to be appreciable~\cite{WFA+99}.
As Eq.~(\ref{eq:solAn}) has a deeper penetration into the bulk,
its degeneracy is lifted first, as can be seen in Fig.~ \ref{cap:edgestateRibbon}~(b).
We may then conclude that edge states do exist in bilayer graphene
ribbons. We expect band gaps to open due to magnetic instabilities
induced by electron-electron interactions, similarly to graphene single
layer \cite{japonese,SCLprl06}. Actually, the edge states we have
found live only on a single sublattice: $A$ or $B$ depending on
the edge they are localized. This kind of localized states favor a
ferromagnetic arrangement along the edge and antiferromagnetic between
edges \cite{Eduardo}, consistent with what is found by first principles
for stacked graphitic strips \cite{LSP+05}. Also half-metallicity
should occur in graphene bilayer nanoribbons as a consequence of edge
states, analogously to the single layer \cite{SCLnat06}.

From the point of view of scanning tunneling microscopy (STM) we notice
that bilayer edge states give rise to different intensities depending
on the ribbon edge. As an example we consider the ribbon shown in
Fig.~\ref{cap:ribbon}, and assume that the STM signal is essentially
proportional to the local density of states of the upper layer. At
the top zigzag edge the STM signal is due to edge states of the \emph{bilayer}
type, the only ones with finite amplitude on the upper layer {[}Eq.~(\ref{eq:solAn})].
On the other hand, at the bottom zigzag edge both \emph{bilayer} and
\emph{monolayer} families have finite amplitude on the upper layer,
and a higher STM intensity is expected therefrom.

%

\begin{figure}[t]
\begin{centering}
\includegraphics[scale=0.35,width=0.9\columnwidth]{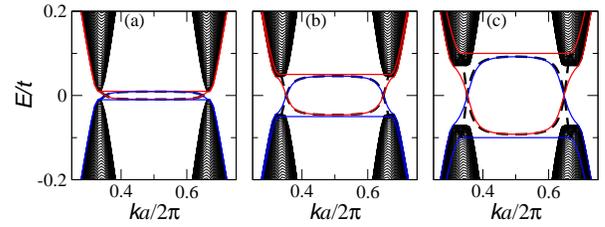}
\par\end{centering}

\caption{\label{cap:edgestateV}(Color online) Energy spectrum for a bilayer
ribbon with zigzag edges for different values of the bias $V$: (a)
$V=t_{\perp}/10$, (b) $V=t_{\perp}/2$, (c) $V=t_{\perp}$. Inter-layer
coupling $t_{\perp}/t=0.2$ and ribbon width $N=400$. The dashed
lines are the analytical result {[}Eq.~(\ref{eq:disp})].}
\end{figure}

\emph{Nanoribbons of bilayer graphene (biased):} It has recently been
shown that the electronic gap of a graphene bilayer can be effectively
controlled externally by applying a gate bias \cite{ENM+06}. We now
consider the case of biased bilayer nanoribbon with zigzag edges,
where the presence of edge states should play a role. The bias gives
rise to an electrostatic potential difference, $V$, between the two
layers. This is parametrized by adding to the Hamiltonian in Eq.~(\ref{eq:H})
the term $\frac{V}{2}\sum_{m,n}(n_{1;m,n}-n_{2;m,n})$, with $n_{i;m,n}=a_{i;m,n}^{\dagger}a_{i;m,n}+b_{i;m,n}^{\dagger}b_{i;m,n}$.
Edge states are strongly affected by the bias. The semi-infinite biased
system has only one edge state given by Eq.~(\ref{eq:solB}), as
the edge state having finite amplitudes at both layers {[}Eq.~(\ref{eq:solAn})]
is no longer an eigenstate. In Fig.~\ref{cap:edgestateV} we show
the band structure of a bilayer ribbon for different values of the
bias. Two partially flat bands for $k$ in the range $2\pi/3\leq ka\leq4\pi/3$
are clearly seen at $E=\pm V/2$. These are bands of edge states localized
at opposite ribbon sides, with finite amplitudes on a single layer
{[}Eq.~(\ref{eq:solB}) and its counterpart for the other edge].
Also evident is the presence of two dispersive bands crossing the
gap. Both the closeness of these dispersive bands to $E\approx\pm V/2$
for $ka\approx\pi$ and their crossing at $E=0$ near the Dirac points
can be understood using perturbation theory in $V/t$. As surface
states living at opposite edges have an exponentially small overlapping,
and those belonging to the same edge are orthogonal, we can treat
the solution given by Eq.~(\ref{eq:solAn}) and its counterpart for
the other edge separately. Starting with Eq.~(\ref{eq:solAn}), the
first order energy shift induced by the applied bias is $E_{k}=V/2(\langle n_{1}^{k}\rangle-\langle n_{2}^{k}\rangle)$,
where $\langle n_{1}^{k}\rangle$ and $\langle n_{2}^{k}\rangle$
give the probability of finding the localized electron in layer 1
and 2, respectively. The value of these quantities is easily obtained
from Eq.~(\ref{eq:solAn}) through a real space summation,%
%
 \begin{align}
\langle n_{1}^{k}\rangle & =\frac{(1-D_{k}^{2})^{2}}{(1-D_{k}^{2})^{2}+t_{\perp}^{2}/t^{2}},\label{eq:n1}\\
\langle n_{2}^{k}\rangle & =\frac{t_{\perp}^{2}/t^{2}}{(1-D_{k}^{2})^{2}+t_{\perp}^{2}/t^{2}}.\label{eq:n2}\end{align}
 The band dispersion is thus given by,%
%
 \begin{equation}
E_{k}^{\pm}=\pm\frac{V}{2}\frac{(1-D_{k}^{2})^{2}-t_{\perp}^{2}/t^{2}}{(1-D_{k}^{2})^{2}+t_{\perp}^{2}/t^{2}},\label{eq:disp}\end{equation}
where the minus sign stands for the band of states localized at the
opposite edge. The result of Eq.~(\ref{eq:disp}) is shown in Fig.~\ref{cap:edgestateV}
as a dashed line which is hardly distinguishable from the numerical
result. Note that for $ka\approx\pi$ we have $D_{k}\rightarrow0$,
so that $E_{k}^{\pm}\approx\pm V/2$. This means that for $ka\approx\pi$
the edge state given by Eq.~(\ref{eq:solAn}) is essentially localized
at layer~1, which is clearly seen from Eqs.~(\ref{eq:n1}) and~(\ref{eq:n2})
as long as $t_{\perp}/t\ll1$. For $1-D_{k}^{2}=t_{\perp}/t$ the
energy shift {[}Eq.~(\ref{eq:disp})] is zero, which leads to band
crossing. For $t_{\perp}\ll t$ we can expand around the Dirac points,
$k_{0}^{\pm}a=2\pi/3,4\pi/3$. If $k=k_{0}+\delta k$, the crossing
takes place for $\delta ka=\pm t_{\perp}/(t\sqrt{3})$, each sign
being assigned to different Dirac points. Note that $\delta k$ is
$V$ independent, and its value compares fairly well with the numerical
results shown in Fig.~\ref{cap:edgestateV}. Indeed, the approximate
result given by Eq.~(\ref{eq:disp}) only fails at the Dirac points,
where the edge states localization Length diverges and their overlap
has to be considered. Increasing the bias makes first order perturbation
theory to break down. We have found numerically that the gap opens
for $V\gtrsim t$.

%

\emph{Conclusions:} We have shown that zero energy edge states do
exist at zigzag edges of bilayer graphene. We have derived an analytic
solution for the wavefunction assuming a semi-infinite system and
a first nearest-neighbor tight-biding model. This analytic solution
defines two types of edge states: monolayer edge states, with finite
amplitude over a single plane; and bilayer edge states, with finite
amplitude over the two planes, and with an enhanced penetration into
the bulk. Edge states are present even in bilayer graphene nanoribbons,
where edge magnetization as well as half-metallicity are expected
to show up in analogy with single layer graphene. We have also shown
the robustness of bilayer graphene edge states to the presence of
an electrostatic potential difference between planes.



~\\
E.V.C. acknowledges the financial support of FCT through Grant No.~SFRH/BD/13182/2003.
E.V.C., N.M.R.P., and J.M.B.L.S. acknowledge financial support from
POCI 2010 via project PTDC/FIS/64404/2006. A.H.C.N was supported through
NSF grant DMR-0343790.

\bibliographystyle{apsrev}

\begin{thebibliography}{15}
\expandafter\ifx\csname natexlab\endcsname\relax\def\natexlab#1{#1}\fi
\expandafter\ifx\csname bibnamefont\endcsname\relax
  \def\bibnamefont#1{#1}\fi
\expandafter\ifx\csname bibfnamefont\endcsname\relax
  \def\bibfnamefont#1{#1}\fi
\expandafter\ifx\csname citenamefont\endcsname\relax
  \def\citenamefont#1{#1}\fi
\expandafter\ifx\csname url\endcsname\relax
  \def\url#1{\texttt{#1}}\fi
\expandafter\ifx\csname urlprefix\endcsname\relax\def\urlprefix{URL }\fi
\providecommand{\bibinfo}[2]{#2}
\providecommand{\eprint}[2][]{\url{#2}}

\bibitem[{\citenamefont{Novoselov et~al.}(2004)\citenamefont{Novoselov, Geim,
  Morozov, Jiang, Zhang, Dubonos, Grigorieva, and Firsov}}]{NGM+04}
\bibinfo{author}{\bibfnamefont{K.}~\bibnamefont{Novoselov}},
  \bibinfo{author}{\bibfnamefont{A.}~\bibnamefont{Geim}},
  \bibinfo{author}{\bibfnamefont{S.}~\bibnamefont{Morozov}},
  \bibinfo{author}{\bibfnamefont{D.}~\bibnamefont{Jiang}},
  \bibinfo{author}{\bibfnamefont{Y.}~\bibnamefont{Zhang}},
  \bibinfo{author}{\bibfnamefont{S.}~\bibnamefont{Dubonos}},
  \bibinfo{author}{\bibfnamefont{I.}~\bibnamefont{Grigorieva}},
  \bibnamefont{and} \bibinfo{author}{\bibfnamefont{A.}~\bibnamefont{Firsov}},
  \bibinfo{journal}{Science} \textbf{\bibinfo{volume}{306}},
  \bibinfo{pages}{666} (\bibinfo{year}{2004}).

\bibitem[{\citenamefont{A.K.Geim and Novoselov}(2007)}]{GN07}
\bibinfo{author}{\bibnamefont{A.K.Geim}} \bibnamefont{and}
  \bibinfo{author}{\bibfnamefont{K.}~\bibnamefont{Novoselov}},
  \bibinfo{journal}{Nature Materials} \textbf{\bibinfo{volume}{6}},
  \bibinfo{pages}{183} (\bibinfo{year}{2007}).

\bibitem[{\citenamefont{Chen et~al.}()\citenamefont{Chen, Lin, Rooks, and
  Avouris}}]{CLR+07}
\bibinfo{author}{\bibfnamefont{Z.}~\bibnamefont{Chen}},
  \bibinfo{author}{\bibfnamefont{Y.-M.} \bibnamefont{Lin}},
  \bibinfo{author}{\bibfnamefont{M.~J.} \bibnamefont{Rooks}}, \bibnamefont{and}
  \bibinfo{author}{\bibfnamefont{P.}~\bibnamefont{Avouris}},
  \bibinfo{note}{cond-mat/0701599}.

\bibitem[{\citenamefont{Han et~al.}(2007)\citenamefont{Han, Oezyilmaz, Zhang,
  and Kim}}]{HOZ+07}
\bibinfo{author}{\bibfnamefont{M.~Y.} \bibnamefont{Han}},
  \bibinfo{author}{\bibfnamefont{B.}~\bibnamefont{Oezyilmaz}},
  \bibinfo{author}{\bibfnamefont{Y.}~\bibnamefont{Zhang}}, \bibnamefont{and}
  \bibinfo{author}{\bibfnamefont{P.}~\bibnamefont{Kim}},
  \bibinfo{journal}{Phys. Rev. Lett} \textbf{\bibinfo{volume}{98}},
  \bibinfo{pages}{206805} (\bibinfo{year}{2007}).

\bibitem[{\citenamefont{Fujita et~al.}(1996)\citenamefont{Fujita, Wakabayashi,
  Nakada, and Kusakabe}}]{japonese}
\bibinfo{author}{\bibfnamefont{M.}~\bibnamefont{Fujita}},
  \bibinfo{author}{\bibfnamefont{K.}~\bibnamefont{Wakabayashi}},
  \bibinfo{author}{\bibfnamefont{K.}~\bibnamefont{Nakada}}, \bibnamefont{and}
  \bibinfo{author}{\bibfnamefont{K.}~\bibnamefont{Kusakabe}},
  \bibinfo{journal}{J. Phys. Soc. Jpn.} \textbf{\bibinfo{volume}{65}},
  \bibinfo{pages}{1920} (\bibinfo{year}{1996}).

\bibitem[{\citenamefont{Nakada et~al.}(1996)\citenamefont{Nakada, Fujita,
  Dresselhaus, and Dresselhaus}}]{Dresselhaus}
\bibinfo{author}{\bibfnamefont{K.}~\bibnamefont{Nakada}},
  \bibinfo{author}{\bibfnamefont{M.}~\bibnamefont{Fujita}},
  \bibinfo{author}{\bibfnamefont{G.}~\bibnamefont{Dresselhaus}},
  \bibnamefont{and} \bibinfo{author}{\bibfnamefont{M.~S.}
  \bibnamefont{Dresselhaus}}, \bibinfo{journal}{Phys. Rev. B}
  \textbf{\bibinfo{volume}{54}}, \bibinfo{pages}{17954} (\bibinfo{year}{1996}).

\bibitem[{\citenamefont{Wakabayashi et~al.}(1999)\citenamefont{Wakabayashi,
  Fujita, Ajiki, and Sigrist}}]{WFA+99}
\bibinfo{author}{\bibfnamefont{K.}~\bibnamefont{Wakabayashi}},
  \bibinfo{author}{\bibfnamefont{M.}~\bibnamefont{Fujita}},
  \bibinfo{author}{\bibfnamefont{H.}~\bibnamefont{Ajiki}}, \bibnamefont{and}
  \bibinfo{author}{\bibfnamefont{M.}~\bibnamefont{Sigrist}},
  \bibinfo{journal}{Phys. Rev. B} \textbf{\bibinfo{volume}{59}},
  \bibinfo{pages}{8271} (\bibinfo{year}{1999}).

\bibitem[{\citenamefont{Peres et~al.}(2006)\citenamefont{Peres, Guinea, and
  Castro~Neto}}]{PGN06}
\bibinfo{author}{\bibfnamefont{N.~M.~R.} \bibnamefont{Peres}},
  \bibinfo{author}{\bibfnamefont{F.}~\bibnamefont{Guinea}}, \bibnamefont{and}
  \bibinfo{author}{\bibfnamefont{A.~H.} \bibnamefont{Castro~Neto}},
  \bibinfo{journal}{Phys. Rev. B} \textbf{\bibinfo{volume}{73}},
  \bibinfo{pages}{125411} (\bibinfo{year}{2006}).

\bibitem[{\citenamefont{Son et~al.}(2006{\natexlab{a}})\citenamefont{Son,
  Cohen, and Louie}}]{SCLprl06}
\bibinfo{author}{\bibfnamefont{Y.-W.} \bibnamefont{Son}},
  \bibinfo{author}{\bibfnamefont{M.~L.} \bibnamefont{Cohen}}, \bibnamefont{and}
  \bibinfo{author}{\bibfnamefont{S.~G.} \bibnamefont{Louie}},
  \bibinfo{journal}{Phys. Rev. Lett.} \textbf{\bibinfo{volume}{97}},
  \bibinfo{pages}{216803} (\bibinfo{year}{2006}{\natexlab{a}}).

\bibitem[{\citenamefont{McCann and Fal'ko}(2006)}]{MF06}
\bibinfo{author}{\bibfnamefont{E.}~\bibnamefont{McCann}} \bibnamefont{and}
  \bibinfo{author}{\bibfnamefont{V.~I.} \bibnamefont{Fal'ko}},
  \bibinfo{journal}{Phys. Rev. Lett.} \textbf{\bibinfo{volume}{96}},
  \bibinfo{pages}{086805} (\bibinfo{year}{2006}).

\bibitem[{\citenamefont{Castro et~al.}({\natexlab{a}})\citenamefont{Castro,
  Novoselov, Morozov, Peres, Lopes~dos Santos, Nilsson, Guinea, Geim, and
  Castro~Neto}}]{ENM+06}
\bibinfo{author}{\bibfnamefont{E.~V.} \bibnamefont{Castro}},
  \bibinfo{author}{\bibfnamefont{K.~S.} \bibnamefont{Novoselov}},
  \bibinfo{author}{\bibfnamefont{S.~V.} \bibnamefont{Morozov}},
  \bibinfo{author}{\bibfnamefont{N.~M.~R.} \bibnamefont{Peres}},
  \bibinfo{author}{\bibfnamefont{J.~M.~B.} \bibnamefont{Lopes~dos Santos}},
  \bibinfo{author}{\bibfnamefont{J.}~\bibnamefont{Nilsson}},
  \bibinfo{author}{\bibfnamefont{F.}~\bibnamefont{Guinea}},
  \bibinfo{author}{\bibfnamefont{A.~K.} \bibnamefont{Geim}}, \bibnamefont{and}
  \bibinfo{author}{\bibfnamefont{A.~H.} \bibnamefont{Castro~Neto}},
  \bibinfo{note}{cond-mat/0611342}.

\bibitem[{\citenamefont{Sasaki et~al.}(2006)\citenamefont{Sasaki, Murakami, and
  Saito}}]{SMS06}
\bibinfo{author}{\bibfnamefont{K.}~\bibnamefont{Sasaki}},
  \bibinfo{author}{\bibfnamefont{S.}~\bibnamefont{Murakami}}, \bibnamefont{and}
  \bibinfo{author}{\bibfnamefont{R.}~\bibnamefont{Saito}},
  \bibinfo{journal}{Appl. Phys. Lett.} \textbf{\bibinfo{volume}{88}},
  \bibinfo{pages}{113110} (\bibinfo{year}{2006}).

\bibitem[{\citenamefont{Castro et~al.}({\natexlab{b}})\citenamefont{Castro,
  Peres, Lopes~dos Santos, Castro~Neto, and Guinea}}]{Eduardo}
\bibinfo{author}{\bibfnamefont{E.~V.} \bibnamefont{Castro}},
  \bibinfo{author}{\bibfnamefont{N.~M.~R.} \bibnamefont{Peres}},
  \bibinfo{author}{\bibfnamefont{J.~M.~B.} \bibnamefont{Lopes~dos Santos}},
  \bibinfo{author}{\bibfnamefont{A.~H.} \bibnamefont{Castro~Neto}},
  \bibnamefont{and} \bibinfo{author}{\bibfnamefont{F.}~\bibnamefont{Guinea}},
  \bibinfo{note}{to be published}.

\bibitem[{\citenamefont{Lee et~al.}(2005)\citenamefont{Lee, Son, Park, Han, and
  Yu}}]{LSP+05}
\bibinfo{author}{\bibfnamefont{H.}~\bibnamefont{Lee}},
  \bibinfo{author}{\bibfnamefont{Y.-W.} \bibnamefont{Son}},
  \bibinfo{author}{\bibfnamefont{N.}~\bibnamefont{Park}},
  \bibinfo{author}{\bibfnamefont{S.}~\bibnamefont{Han}}, \bibnamefont{and}
  \bibinfo{author}{\bibfnamefont{J.}~\bibnamefont{Yu}}, \bibinfo{journal}{Phys.
  Rev. B} \textbf{\bibinfo{volume}{72}}, \bibinfo{pages}{174431}
  (\bibinfo{year}{2005}).

\bibitem[{\citenamefont{Son et~al.}(2006{\natexlab{b}})\citenamefont{Son,
  Cohen, and Louie}}]{SCLnat06}
\bibinfo{author}{\bibfnamefont{Y.-W.} \bibnamefont{Son}},
  \bibinfo{author}{\bibfnamefont{M.~L.} \bibnamefont{Cohen}}, \bibnamefont{and}
  \bibinfo{author}{\bibfnamefont{S.~G.} \bibnamefont{Louie}},
  \bibinfo{journal}{Nature} \textbf{\bibinfo{volume}{444}},
  \bibinfo{pages}{347} (\bibinfo{year}{2006}{\natexlab{b}}).

\end{thebibliography}

\end{document}